\documentstyle[12pt]{article}
\begin{document}
\baselineskip=13pt
\begin{center} 
{\bf Ising Model on A Fibonacci Lattice with Magnetic Field}
\end{center}
\vspace{1.0cm} 
\begin{center}
{\bf Susanta Bhattacharya}$^a$ and {\bf Samir K. Paul}$^b$\\
$^a$ Ramsaday College\\
Amta,Howrah,West Bengal,India\\

$^b$ S. N. Bose National Centre For Basic Sciences\\
Block-JD, Sector-III, Salt Lake\\
Calcutta-700091,  India
\end{center}

\vspace{0.5cm}

\noindent        {\bf Abstract}

   We present a general procedure for calculating the partition function of an  Ising Model on a one dimensional  Fibonacci lattice in presence of magnetic     field .This partition function can be written as a sum of partition functions of
 usual open Ising chains in presence of magnetic field with coeffecients
having Fibonacci symmetries.An 
exact expression for the partition function of the usual open open Ising
Model is found.We observe that "Mirror Symmetry"        
is a characteristic property of all Fibonacci generations . Further $nth$ and $(n+6)th$
generations have the same topology.We have also established a recurrence        relation among  partition functions of different Fibonacci generations from     $nth$ to $(n+6)th$.


\vspace{0.5cm}

a)  susanta1005@rediffmail.com\\
b)  smr@boson.bose.res.in

\vspace{0.5cm}

\noindent        {\bf Introduction}

      Studying Ising model on a Fibonacci chain in presence of magnetic field
is interesting in its own right since no proper analytical procedure exists
for evaluating the partition function.However scaling forms of thermodynamic
functions for such system have been studied using renormalization group technique
[1] through one step decimation.Here we reformulate the expression for the     partition function by breaking the transfer matrix into two particular noncommuting matrices . This formulation 
enables us to calculate the partition function for the usual open Ising chain.
The result is quite nontrivial in contrast to the expression for the partition  function of the closed one[2].
The same
formulation helps us to express the partition function of Ising model on Fibonacci chain in presence of magnetic field as a sum  of partition  functions of usual Ising open chains with coefficients containing Fibonacci symmetry.We have also studied some symmetry properties of the Fibonacci chain.Using a special symmetry property    ("Mirror Symmetry") and the usual trace map relation we have established a 
recurrence relation among the partition functions of different Fibonacci
generations.This includes all the partition functions starting from $nth$ up to
$(n+6)th$ generations.We observe that mirror symmetry is a characteristic property of each Fibonacci generation with $nth$ and $(n+6)th$ generations having  
same topology .

\vspace{2.0cm}

\noindent {\bf {Exact partition function for open Ising chain with magnetic field}}

\vspace{0.5cm}

  The one dimensional Ising model consists of a chain of N spins ${S_i}={\pm 1}$

 ;$i =1,2,.....,N$ with nearest neighbour interactions ${\epsilon_{i,i+1}}$. The

 Hamiltonian is given by:

\begin{equation}
{\mathcal{H}} = - {\sum_{i=1}^{N-1}}{\epsilon_{i,i+1}} {S_i} {S_{i+1}} - H{\sum_{i=1}^{N}}{S_i}
\end{equation}

For a uniform lattice ${\epsilon_{i,i+1}} = {\epsilon}$,the partition function   is given by: 
 
\begin{equation}
{Z_N^o}(T,H) = {\sum_{{S_1},{S_2},....,{S_N}= - 1}^{+1}} f({S_1},{S_2})
f({S_2},{S_3})......f({S_{N-1}},{S_N}) {f_0}({S_N},{S_1})
\end{equation}

with
$f({S_i},{S_{i+1}}) =exp[{\beta}{\epsilon}{S_i}{S_{i+1}} + {\frac{1}{2}}
\noindent       {\beta}H({S_i}+{S_{i+1}})]$; 
${f_0}({S_N},{S_1}) = {{[f({S_N},{S_1})]}_{\epsilon =0}}$.                      \noindent   Here the superscript $o$ stands for the chain with open
boundary condition.Therefore the partition function (2) can be written in terms
of transfer matrix as: 

\begin{equation}
{Z_N^o}(T,H) = Tr {P^{N-1}}{P_0}
\end{equation}

where
\begin{equation}
P = {\sqrt{r}}(1+{\frac{\lambda}{r}}){{\sigma}_1} = {\sqrt{r}}{{\sigma}_1}(1+
{\frac{{\lambda}^T}{r}})
\end{equation}

\begin{equation}
{P_0}={[P]_{\epsilon =0}}=(1+{\lambda}){\sigma_1}={\sigma_1}(1+{\lambda^T})
\end{equation}
 
with 
$r = exp(-2{\beta}{\epsilon})$ , 
$\lambda = \left(\matrix{0 & e^{\beta H}\cr e^{-\beta H} & 0}\right),$
${\lambda^T} = \left(\matrix{0 & e^{-\beta H}\cr e^{\beta H} & 0}\right)$
and ${\sigma_1} = \left(\matrix{0 & 1 \cr 1 & 0}\right).$ 

\vspace{0.2cm}

 The general formula of the partition function for even and odd number

of spins (i.e.,odd and even number of bonds) can be derived by using equations 

(3) and (4) as :

\begin{equation}
{Z_{2N}^o}(T,H) = {r^{N-{\frac{1}{2}}}}Tr(1+{x_1})(1+{x_2})....(1+{x_{2N-1}})
(1+{\lambda^T})
\end{equation}

 and 

\begin{equation}
{Z_{2N+1}^o}(T,H) = {r^N}Tr(1+{x_1})(1+{x_2})......(1+{x_{2N}})(1+\lambda )
{\sigma_1}
\end{equation}
 
where 

\begin{equation}
{x_{2i+1}} = {\frac{\lambda}{r}},     {x_{2i}} = {\frac{\lambda^T}{r}}; i=integer     
\end{equation}

The above equations show that $\lambda , {\lambda^T}$ are the signatures for
the transfer matrices corresponding to bonds in odd and even positions.In the
case of a chain with closed boundary condition the last factor in eqn.(2) is
$f({S_N},{S_1})$ and consequently the partition function takes the form: 

\begin{equation}
{Z_{2N}^c}(T,H) = {r^N} Tr(1+{x_1})(1+{x_2})......(1+{x_{2N}})
\end{equation} 

\begin{equation}
{Z_{2N+1}^c} = {r^{N+{\frac{1}{2}}}}Tr(1+{x_1})(1+{x_2}).....(1+{x_{2N+1}})
{\sigma_1}
\end{equation} 

Here the superscript $c$  indicates closed chain.One can show by elementary calculation that eqns.(9) and (10) reduce to the well known form[3] 

\begin{equation}
{Z_N^c}(T,H) = {\lambda_{+}^N} + {\lambda_{-}^N}
\end{equation} 

where 

\begin{equation}
{\lambda_{\pm}} = {r^{-\frac{1}{2}}}[{\cosh{(\beta H)}} \pm {\sqrt{({{\sinh}^2}
{(\beta H)} + {r^2})}}]
\end{equation} 

are the eigenvalues of the transfer matrix $P$.The expression for the partition  function in the case of an open chain with even number of spins can be derived from eqn.(6) as follows: 

\begin{eqnarray} 
{Z_{2N}^o}(T,H) 
 = {r^{N-{\frac{1}{2}}}}Tr(1+{x_1})(1+{x_2}).....(1+{x_{2N-1}})(1+\lambda^T)\nonumber\\
 = {\sqrt{r}}{Z_{2N}^c}(T,H) + {r^{N-{\frac{1}{2}}}}(1-r)Tr(1+{x_1})(1+{x_2})
.......(1+{x_{2N-1}})\nonumber\\
 = {\sqrt{r}}{Z_{2N}^c}(T,H) + {\sqrt{r}}(1-r){Z_{2(N-1)}^c}(T,H) + 
{r^{N-{\frac{1}{2}}}}(1-r)\nonumber\\
\times Tr (1+{x_1})(1+{x_2}).......(1+{x_{2N-2}}){x_{2N-1}}
\end{eqnarray}

The last term in the above expression can be written in terms of the eigenvalues of the transfer matrix $P$ viz.${\lambda_\pm}$. By following the method of induction :

\begin{eqnarray} 
{r^{N-{\frac{1}{2}}}}(1-r)Tr(1+{x_1})(1+{x_2})......(1+{x_{2N-2}}){x_{2N-1}}
\nonumber\\
= (1-r){r^{N-{\frac{1}{2}}}}{\frac{4}{r^2}}{\cosh^2}{(\beta H)}
{\sum_{i=0}^{N-2}}{\bigg( {\frac{{\lambda_+}^2}{r}} \bigg)}^{N-2-i}   
 {\bigg( {\frac{{\lambda_-}^2}{r}} \bigg)}^{i}\nonumber\\                                   
=4(1-r){r^{-\frac{1}{2}}}{\cosh^2}(\beta H)\frac{{{\lambda_+}^{2(N-1)}}-{{\lambda_-}^{2(N-1)}}}{{{\lambda_+}^2}-
{{\lambda_-}^2}}
\end{eqnarray} 

So eqn.(13) becomes : 

\begin{eqnarray}
{Z_{2N}^o}(T,H)& = &{\sqrt{r}}{Z_{2N}^c}(T,H)+{\sqrt{r}}(1-r){Z^c_{2(N-1)}}(T,H)\nonumber\\                                                                     &  &+ 4(1-r){r^{-\frac{1}{2}}}{\cosh^2}
{(\beta H)}
\times {\frac{{{\lambda_+}^{2(N-1)}}-{{\lambda_-}^{2(N-1)}}}{{{\lambda_+}^2}
- {{\lambda_-}^2}}}
\end{eqnarray}
 
Similarly the expression (7) for the open chain partition function with odd
number of spins takes the form:

\begin{eqnarray}
{Z_{2N+1}^o}(T,H) = {\sqrt{r}}{Z_{2N+1}^c}+2(1-r){\cosh{(\beta H)}}\nonumber\\
\times \frac{{{\lambda_+}^{2N}}-{{\lambda_-}^{2N}}}{{{\lambda_+}^2}-{{\lambda_+}^2}}
\end{eqnarray} 

Exact expressions for the thermodynamic functions can be calculated by well
known methods [3].

\vspace{0.5cm}

\noindent {\bf {Ising model on Fibonacci chain}}

\vspace{0.2cm}

  A Fibonacci chain can be inflated by two bonds $L(large)$ and $S(small)$ by the
inflation rule$L \longrightarrow LS,S \longrightarrow L$.The chain can be
represented by the sequence: 

\begin{equation}
L \longrightarrow LS \longrightarrow LSL \longrightarrow LSLLS \longrightarrow 
 LSLLSLSL \longrightarrow .....
\end{equation}

In this case the interaction strengths in the Hamiltonian (1) ${\epsilon_{i,i+1}}=\epsilon$ for long bonds and ${\epsilon_{i,i+1}}={\bar \epsilon}$ for the short ones where the bonds are arranged according to the Fibonacci sequence (17).The corresponding partition function of the $nth$ generation Fibonacci chain having 
$N$ spins with $N-1$ bonds is given by:

\begin{equation}
{Z_N^o}(F)=Tr P{\bar P}PP{\bar P}.....{P_0}
\end{equation}

where for long bonds the transfer matrix $P$ is given by eqn.(4) and for short 
bonds the transfer matrix  ${\bar P}$ is given by eqn.(4) with $r$ replaced by
${\bar r}= {r|}_{\epsilon = {\bar \epsilon}}$.Henceforth ${Z_N^o}(F)$ and ${Z_N^o}(I)$ will represent partition functions for  Ising models on an open Fibonacci
chain and on an open regular lattice respectively.The expressions for the partition functions with odd and even number of bonds take the same forms as shown in eqns.(6) and (7) with $x_i$'s given in eqn.(8) for long bonds whereas for short bonds
we replace $r$ by ${\bar r}$ in eqn.(8).The explicit expressions for the partition functions for open and closed chains are: 

\begin{equation}
{Z_{2N}^o}(F)={r^{\frac{N_L}{2}}}{{\bar r}^{\frac{N_S}{2}}}Tr (1+{x_1})(1+{x_2})......(1+{x_{2N-1}})({1+\lambda^T})
\end{equation}
    
\begin{equation}
{Z_{2N+1}^o}(F)={r^{\frac{N_L}{2}}}{{\bar r}^{\frac{N_S}{2}}}Tr (1+{x_1})(1+{x_2})......(1+x_{2N})({1+\lambda}){\sigma_1}
\end{equation}

Similarly for closed chain eqs. (18),(19) and (20) take the following forms:

\begin{equation}
{Z_N^c}(F)= Tr P{\bar P}PP{\bar P}.......P
\end{equation}

\begin{equation}
{Z_{2N}^c}(F)={r^{\frac{N_L}{2}}}{{\bar r}^{\frac{N_S}{2}}}Tr (1+{x_1})(1+{x_2})......(1+{x_{2N-1}})(1+{x_{2N}})
\end{equation}

\begin{equation}
{Z_{2N+1}^c}(F)={r^{\frac{N_L}{2}}}{{\bar r}^{\frac{N_S}{2}}}Tr(1+{x_1})(1+{x_2}).......(1+{x_{2N}})(1+{x_{2N+1}}){\sigma_1}
\end{equation}

where $N_L$,$N_S$ are number of long and short bonds in a particular sequence.

\vspace{0.1cm} 
 
Now ${\bar P}$ is related to P through the following equation:

\begin{equation}
{\bar P}={\sqrt{\bar r}}(1-{\frac{r}{\bar r}}){\sigma_1}+{\sqrt{{\frac{r}{\bar r}}}}   P
\end{equation}

Using eqn.(24) in eqns.(19) and (20) the Fibonacci partition function for any generation can be written in terms of open Ising partition functions as follows:

\begin{equation}
{Z_{2N}^o}(F)={h_0}({\epsilon,{\bar \epsilon}})+{\sum_{i=1}^N}{h_{2i}}({\epsilon, {\bar \epsilon}}){Z_{2i}^o}(I)
\end{equation}

\begin{equation}
{Z_{2N-1}^o}(F)={l_0}({\epsilon,{\bar \epsilon}})+{\sum_{i=1}^N}{l_{2i-1}}({\epsilon,{\bar \epsilon}}){Z_{2i-1}^o}(I)
\end{equation}

where ${Z_{2i}^o}(I)$ and ${Z_{2i-1}^o}(I)$ are given by eqns.(15) and (16) respectively.We observe that the quasiperiodic nature of the Fibonacci chain is encoded in the functions $h(\epsilon,{\bar \epsilon})$ and $l(\epsilon,{\bar \epsilon})$.Though for small generations these functions can be derived exactly still we could not find out their general forms. 

\vspace{0.1cm}

To circumvent this difficulty we study the recurrence  relations among the partition functions of different Fibonacci generations.A servey of different Fibonacci generations depicted by eqn.(17) shows a symmetric pattern in terms of the number of bonds ,viz.,

\begin{eqnarray}
{P_1} : P (odd)\nonumber\\
{P_2} : P{\bar P}(even)\nonumber\\
{P_3} : P{\bar P}P(odd)\nonumber\\
{P_4} : P{\bar P}PP{\bar P}(odd)\nonumber\\
{P_5} : P{\bar P}PP{\bar P}P{\bar P}P(even)\nonumber\\
{P_6} : P{\bar P}PP{\bar P}P{\bar P}PP{\bar P}PP{\bar P}(odd)
\end{eqnarray}
and so on.

\vspace{0.5cm}

\noindent {\bf {Recurrence relation among partition functions}}

\vspace{0.2cm}

    Let $P_{n-2}$ be the $(n-2)th$  Fibonacci generation with even number of bonds.This automatically ensures that the previous as well as the next two consecutive generations will have odd number of bonds . The recurrence  relation for   Fibonacci generations is given by:

\begin{equation}
{P_n}={P_{n-1}}{P_{n-2}}
\end{equation}

Now adding a term ${D_{n-2}}{P_{n-3}}$ in the above equation gives

\begin{equation}
{P_n} + {D_{n-2}}{P_{n-3}} = {P_{n-1}}{P_{n-2}} + {D_{n-2}}{P_{n-3}}
\end{equation}

where ${D_{n-2}}=Det({P_{n-2}})$.The following operations are applied            sequentially on eqn.(29): 

\vspace{0.1cm}

     Substitute ${{P_{n-2}}^{-1}}{P_{n-1}}$ in place of ${P_{n-3}}$ on
the right hand side and finally use Cayley-Hamilton theorem to get the usual trace map relation on the Fibonacci lattice:

\begin{equation}
Tr {P_n} = Tr {P_{n-1}} Tr {P_{n-2}} - {D_{n-2}}Tr {P_{n-3}}
\end{equation}
 
 The above equation will be necessary for calculating recurrence relation among
different Fibonacci generations.
For this purpose  we must      understand symmetry properties of Fibonacci chain.Inspecting different generations of the Fibonacci chain it reveals that if the  total number of bonds $N$ of a particular generation is odd then there is a 
 mirror reflection symmetry  arround the ${\bigg( {\frac{N-1}{2}} \bigg)
}th$ bond ;except the last two bonds.If the special bond arround which mirror symmetry
 occurs is a short(long) one the Fibonacci generation will have equal number of  odd and even short(long) bonds . However if the total number of bonds $N$ is
even,the mirror reflection symmetry is arround a cluster of two successive long  bonds at the ${\bigg( {\frac{N}{2}} \bigg)}th$ and ${\bigg( {\frac{N-2}{2}} \bigg)}th$ positions of the chain . 
So "Mirror reflection symmetry"  is a characteristic property of a Fibonacci
chain.

\vspace{0.1cm}

 The $nth$ and $(n\pm 3)th$ generations have the  mirror reflection symmetry property arround the same kind of bond with last two bonds interchanged , while the $(n\pm 6)th$ generations are topologically same as the $nth$ one.

\vspace{0.2cm}

   Using recurrence relation (28) we can write

\begin{equation}
{D_{n-2}}{P_{n-3}}={D_{n-2}}{{P_{n-2}}^{-1}}{P_{n-1}}
\end{equation}

Using Cayley-Hamilton theorem on the right hand side of eq.(31) we get:

\begin{equation}
{D_{n-2}}{P_{n-3}}=(Tr{P_{n-2}}){P_{n-1}}-{P_{n-2}}{P_{n-1}}
\end{equation}

Multiplying eq.(32) by P from the right and taking trace we obtain:

\begin{equation}
{D_{n-2}}{Z^c_{n-3}}=(Tr{P_{n-2}}){Z^c_{n-1}}-Tr({P_{n-2}}{P_{n-1}}P)
\end{equation}

In a  similar fashion we obtain :

\begin{equation}
{D_{n-2}}{Z^o_{n-3}}=(Tr{P_{n-2}}){Z^o_{n-1}}-Tr({P_{n-2}}{P_{n-1}}{P_0})
\end{equation} 

The expression ${P_{n-2}}{P_{n-1}}$ in eqns.(33) and (34) is similar to 
${P_n}={P_{n-1}}{P_{n-2}}$ with last two bonds interchanged,i.e.,both of them
have the same mirror symmetric part $\Omega_n$.Therefore eqns.(33) and (34) can
 be written as: 

\begin{equation}
{D_{n-2}}{Z^c_{n-3}}={Z^c_{n-1}}(Tr{P_{n-2}})-Tr({\Omega_n}P{\bar P}P)
\end{equation}

\begin{equation}
{D_{n-2}}{Z^o_{n-3}}={Z^o_{n-1}}(Tr{P_{n-2}})-Tr({\Omega_n}P{\bar P}{P_0})
\end{equation}

The transfer matrix has the property that $P={P^T}$ and ${\bar P}={{\bar P}^T}$
.If such transfer matrices are arranged in a mirror symmetric fashion then the resulting matrix $({\Omega_n})$ will have the following properties:

\vspace{0.2cm}

i) Off diagonal elements are same i.e.,${({\omega_n})_{12}}={({\omega_n})_{21}}$

\vspace{0.1cm}

ii)Diagonal elements are not same but satisfy the condition:

\vspace{0.1cm}

${({\omega_n})_{11}}(p,q)={({\omega_n})_{22}}(q,p)$;                            where $p={e^{{\beta}H}}$,$q={e^{-{\beta}H}}$. 

\vspace{0.1cm}

Thus the matrix ${\Omega_n}$ in
eqns. (34) and (35) is of the form: 

\vspace{0.2cm}

${\Omega_n}=\left(\matrix {{({\omega_n})_{11}} & {({\omega_n})_{12}}\cr 
{({\omega_n})_{21}} & {({\omega_n})_{22}}}\right)$ 

\vspace{0.2cm}

Eqns.(35) and (36) can be written explictly in the following way:

\begin{eqnarray}
{D_{n-2}}{Z^c_{n-3}}& = & {Z^c_{n-1}}(Tr{P_{n-2}})
-r{\sqrt{\bar r}}[{({\omega_n})_{11}}(y+{\frac{pu}{r}})+{({\omega_n})_{12}}(u+v+{\frac{px+qy}{r}})\nonumber\\
+{({\omega_n})_{22}}(x+{\frac{qv}{r}})]
\end{eqnarray}

\begin{eqnarray}
{D_{n-2}}{Z^o_{n-3}}& = & {Z^o_{n-1}}(Tr{P_{n-2}})
-{\sqrt{r{\bar r}}}[{({\omega_n})_{11}}(y+pu)+{({\omega_n})_{12}}(u+v+px+qy)\nonumber\\
+{({\omega_n})_{22}}(x+qv)]
\end{eqnarray}

where we have used

\vspace{0.1cm}

\hspace{0.5cm}

$P{\bar P}={\sqrt{r{\bar r}}}\left(\matrix {u & y\cr x & v}\right)$

\vspace{0.1cm}

with 

\vspace{0.2cm} 
 
\hspace{0.5cm}

$x={\frac{p}{\bar r}}+{\frac{q}{r}}$,
$y={\frac{p}{r}}+{\frac{q}{\bar r}}$,$u=1+{\frac{p^2}{r{\bar r}}}$ and
$v=1+{\frac{q^2}{r{\bar r}}}$. 

\vspace{0.2cm}

Eliminating $Tr{P_{n-2}}$ from eqns.(37) and (38) we have:

\begin{eqnarray}
{\frac{y{V_{n-1}}+pu{{V^\prime}_{n-1}}}{(u+v){V_{n-1}}+(px+qy){{V^\prime}_{n-1}}}} {({\omega_n})_{11}}
+{\frac{x{V_{n-1}}+qv{{V^\prime}_{n-1}}}{(u+v){V_{n-1}}+(px+qy){{V^\prime}_{n-1}}}} {({\omega_n})_{22}}
+{({\omega_n})_{12}}\nonumber\\
={\frac{D_{n-2}}{r{\sqrt{\bar r}}}}\times {\frac{{Z_{n-1}^o}{Z_{n-3}^c}-{Z_{n-1}^c}{Z_{n-3}^o}}{(u+v){V_{n-1}}+(px+qy){{V^\prime}_{n-1}}}}
\end{eqnarray}

where 

\vspace{0.2cm}

${V_{n-1}}={\frac{Z_{n-1}^c}{\sqrt{r}}}-{Z_{n-1}^o}$ and
${{V^\prime}_{n-1}}={\frac{Z_{n-1}^c}{\sqrt{r}}}-{\frac{Z_{n-1}^o}{r}}$.

\vspace{0.2cm}

To solve for ${({\omega_n})_{11}}$,${({\omega_n})_{12}}$ and ${({\omega_n})_{22}}$ we need another two equations.These equations are obtained from the usual formulae: 

\vspace{0.2cm}

${Z_n^c}=Tr({\Omega_n}{\bar P}PP)$,   ${Z_n^o}=Tr({\Omega_n}{\bar P}P{P_0})$. 

\vspace{0.2cm}

The explicit forms of these two relations are:

\begin{eqnarray}
{\frac{x+{\frac{pu}{r}}}{u+v+{\frac{qx+py}{r}}}} {({\omega_n})_{11}}
+{\frac{y+{\frac{qv}{r}}}{u+v+{\frac{qx+py}{r}}}} {({\omega_n})_{22}}
+{({\omega_n})_{12}}\nonumber\\
 = {\frac{Z_n^c}{r{\sqrt{\bar r}}}} {\frac{1}{u+v+{\frac{qx+py}{r}}}}
\end{eqnarray}

\begin{eqnarray}
{\frac{x+pu}{u+v+qx+py}}  {({\omega_n})_{11}}
+{\frac{y+qv}{u+v+qx+py}} {({\omega_n})_{22}}+{({\omega_n})_{12}}\nonumber\\
 = {\frac{Z_n^o}{{\sqrt{r{\bar r}}}}} {\frac{1}{u+v+qx+py}}
\end{eqnarray}

By elementary calculation one obtains:

\begin{eqnarray}
{({\omega_n})_{11}}(x,y)=
{\frac{{\Gamma_{n-1}}-{{{\Gamma}^\prime}_{n-1}}}{{\Gamma_{n-1}}{{{\Lambda}^\prime}_{n-1}}-{{{\Gamma}^\prime}_{n-1}}{\Lambda_{n-1}}}}\times {\frac{D_{n-2}}{r\sqrt{\bar r}}}\times {\Delta_{n-1}}\nonumber\\
-{\frac{1}{\sqrt{r{\bar r}}}}{\frac{1}{{\Gamma_{n-1}}{{{\Lambda}^\prime}_{n-1}}-{{{\Gamma}^\prime}_{n-1}}{\Lambda_{n-1}}}}\times ({\Gamma_{n-1}}{Z_n^o}{K^\prime}-{{\Gamma}^\prime}{\frac{Z_n^c}{\sqrt{r}}}K)
\end{eqnarray}

\begin{eqnarray}
{({\omega_n})_{22}}(x,y)=-{\frac{{\Lambda_{n-1}}-{{{\Lambda}^\prime}_{n-1}}}{{\Gamma_{n-1}}{{{\Lambda}^\prime}_{n-1}}-{{{\Gamma}^\prime}_{n-1}}{\Lambda_{n-1}}}}\times {\frac{D_{n-2}}{r\sqrt{\bar r}}}\times {\Delta_{n-1}}\nonumber\\
+{\frac{1}{\sqrt{r{\bar r}}}}{\frac{1}{{\Gamma_{n-1}}{{{\Lambda}^\prime}_{n-1}}-{{{\Gamma}^\prime}_{n-1}}{\Lambda_{n-1}}}}\times ({\Gamma_{n-1}}{Z_n^o}{K^\prime}-{{\Gamma}^\prime}{\frac{Z_n^c}{\sqrt{r}}}K)
\end{eqnarray}

and 

\begin{equation}
{({\omega_n})_{12}}(x,y)={\frac{1}{(qx+py)(1-{\frac{1}{r}})}}\bigg[ -{\frac{1}{\sqrt{r{\bar r}}}}\times {V_n}-{(1-{\frac{1}{r}})}\bigg( pu{({\omega_n})_{11}}+qv{({\omega_n})_{22}} \bigg) \bigg]
\end{equation}

where

\begin{equation}
{\Lambda_{n-1}}(x,y)={\frac{y{V_{n-1}}+pu{{V^\prime}_{n-1}}}{(u+v){V_{n-1}}+(px+qy){{V^\prime}_{n-1}}}}-{\frac{x+{\frac{pu}{r}}}{u+v+{\frac{qx+py}{r}}}} 
\end{equation}

\begin{equation}
{{{\Lambda}^\prime}_{n-1}}(x,y)={\frac{y{V_{n-1}}+pu{{V^\prime}_{n-1}}}{(u+v){V_{n-1}}+(px+qy){{V^\prime}_{n-1}}}}-{\frac{x+pu}{u+v+qx+py}}
\end{equation}

\begin{equation}
{\Gamma_{n-1}}(x,y)=\frac{x{V_{n-1}}+qv{{V^\prime}_{n-1}}}{(u+v){V_{n-1}}+(px+qy){{V^\prime}_{n-1}}}-\frac{y+{\frac{qv}{r}}}{u+v+{\frac{qx+py}{r}}} 
\end{equation}

\begin{equation}
{{{\Gamma}^\prime}_{n-1}}(x,y)= \frac{x{V_{n-1}}+qv{{V^\prime}_{n-1}}}{(u+v){V_{n-1}}+(px+qy){{V^\prime}_{n-1}}}-\frac{y+qv}{u+v+qx+py} 
\end{equation}

\begin{equation}
{\Delta_{n-1}}(x,y)={\frac{{Z_{n-1}^o}{Z_{n-3}^c}-{Z_{n-1}^c}{Z_{n-3}^o}}{(u+v){V_{n-1}}+(px+qy){{V^\prime}_{n-1}}}} 
\end{equation}

\begin{equation}
K(x,y)={\frac{1}{u+v+{\frac{qx+py}{r}}}}
\end{equation}

\begin{equation}
{K^\prime}(x,y)={\frac{1}{u+v+qx+py}} 
\end{equation}

Eliminating $D_{n-2}$ from eqns.(37) and (38) we obtain:

\begin{eqnarray}
Tr{P_{n-2}}={\frac{r{\sqrt {\bar r}}\times {V_{n-3}}}{{Z^o_{n-1}}{Z^c_{n-3}}-{Z^c_{n-1}}{Z^o_{n-3}}}} \bigg[ \bigg( {\alpha_{xy}}+{\beta_{xy}}{\frac{{V^\prime}_{n-3}}{V_{n-3}}}\bigg){({\omega_n})_{11}}(x,y)\nonumber\\
+\bigg( {{\alpha^\prime}_{xy}}+{{\beta^\prime}_{xy}} {\frac{{V^\prime}_{n-3}}{V_{n-3}}} \bigg){({\omega_n})_{22}}(x,y)\nonumber\\
-{\frac{1}{{\sqrt{r\bar r}}(1-{\frac{1}{r}})(qx+py)}}\bigg( (u+v)+(px+qy){\frac{{V^\prime}_{n-3}}{V_{n-3}}} \bigg) {V_n} \bigg]
\end{eqnarray}

where 

\vspace{0.2cm}

${\alpha_{xy}}=y-{\frac{u+v}{py+qx}}\times pu$, \hspace{0.5cm}          ${\beta_{xy}}={\frac{(x-y)(q-p)}{py+qx}}\times pu$

\vspace{0.2cm}

${{\alpha^\prime}_{xy}}=x-{\frac{u+v}{py+qx}}\times qv$,  \hspace{0.5cm} 
${{\beta^\prime}_{xy}}={\frac{(x-y)(q-p)}{py+qx}}\times qv$  

\vspace{0.2cm}
 
In general $n$th and $(n\pm 2)$th generations have same arrangement of the last two bonds appart from their respective mirror symmetric parts.That is why $Tr{P_n}$ and $Tr{P_{n\pm 2}}$ will have similar expressions.Since we have assumed
${P_n}={\Omega_n}{\bar P}P$ it follows from the expression (27) that ${P_{n-3}}={\Omega_{n-3}}P{\bar P}$.Therefore proceeding in a similar way as above we get:
 
\begin{eqnarray}
Tr{P_{n-3}}={\frac{r{\sqrt{\bar r}}\times {V_{n-4}}}{{Z^o_{n-2}}{Z^c_{n-4}}-{Z^c_{n-2}}{Z^o_{n-4}}}}\bigg[ \bigg( {\alpha_{yx}}+{\beta_{yx}}{\frac{{V^\prime}_{n-4}}{V_{n-4}}}\bigg){({{\omega}_{n-1}})_{11}}(y,x)\nonumber\\
+\bigg( {{\alpha^\prime}_{yx}}+{{\beta^\prime}_{yx}}{\frac{{V^\prime}_{n-4}}{V_{n-4}}} \bigg) {({{\omega}_{n-1}})_{11}}(y,x)\nonumber\\
-{\frac{1}{{\sqrt{r\bar r}}(1-{\frac{1}{r}})(px+qy)}}\bigg( (u+v)+(qx+py){\frac{{V^\prime}_{n-4}}{V_{n-4}}} \bigg) {V_{n-1}} \bigg]
\end{eqnarray}

Using eqns. (52) and (53) and similar expressions for $Tr{P_n}$ , $Tr{P_{n-1}}$ in the trace map relation (30) we obtain the following recurrence  relation among partition functions of different Fibonacci generations as:

\begin{eqnarray}
{\frac{V_{n-1}}{{Z^o_{n+1}}{Z^c_{n-1}}-{Z^c_{n+1}}{Z^o_{n-1}}}} \bigg[ \bigg( {\alpha_{xy}}+{\beta_{xy}} {\frac{{V^\prime}_{n-1}}{V_{n-1}}} \bigg) {({\omega_{n+2}})_{11}}(x,y)+\bigg( {{\alpha^\prime}_{xy}}+{{\beta^\prime}_{xy}}{\frac{{V^\prime}_{n-1}}{V_{n-1}}} \bigg) {({\omega_{n+2}})_{22}}(x,y)\nonumber\\
-{\frac{1}{{\sqrt{r\bar r}}(1-{\frac{1}{r}})(qx+py)}}\bigg( u+v+(px+qy){\frac{{V^\prime}_{n-1}}{V_{n-1}}} \bigg){V_{n+2}} \bigg] ={\frac{r{\sqrt{\bar r}}\times {V_{n-2}}}{{Z^o_n}{Z^c_{n-2}}-{Z^c_n}{Z^o_{n-2}}}}\nonumber\\
\times {\frac{V_{n-3}}{{Z^o_{n-1}}{Z^c_{n-3}}-{Z^c_{n-1}}{Z^o_{n-3}}}} \bigg[ \bigg( {\alpha_{yx}}+{\beta_{yx}}{\frac{{V^\prime}_{n-2}}{V_{n-2}}} \bigg){({\omega_{n+1}})_{11}}(y,x)+ \bigg( {{\alpha^\prime}_{yx}}+{{\beta^\prime}_{yx}} {\frac{{V^\prime}_{n-2}}{V_{n-2}}} \bigg) {({\omega_{n+1}})_{22}}(y,x)\nonumber\\
-{\frac{1}{{\sqrt{r\bar r}}(1-{\frac{1}{r}})(px+qy)}}\bigg( u+v+(qx+py) {\frac{{V^\prime}_{n-2}}{V_{n-2}}} \bigg){V_{n+1}} \bigg]   \times \bigg[ \bigg( {\alpha_{xy}}+{\beta_{xy}} {\frac{{V^\prime}_{n-3}}{V_{n-3}}} \bigg) {({\omega_n})_{11}}(x,y)\nonumber\\
+ \bigg( {{\alpha^\prime}_{xy}}+{{\beta^\prime}_{xy}}{\frac{{V^\prime}_{n-3}}{V_{n-3}}} \bigg) {({\omega_n})_{22}}(x,y)- {\frac{1}{{\sqrt{r\bar r}}(1-{\frac{1}{r}})(qx+py)}}\bigg( u+v+(px+qy){\frac{{V^\prime}_{n-3}}{V_{n-3}}} \bigg) {V_n} \bigg]\nonumber\\
-{\frac{{D_{n-2}}{V_{n-4}}}{{Z^o_{n-2}}{Z^c_{n-4}}-{Z^c_{n-2}}{Z^o_{n-4}}}} \bigg[ \bigg( {\alpha_{yx}}+{\beta_{yx}} {\frac{{V^\prime}_{n-4}}{V_{n-4}}} \bigg) {({\omega_{n-1}})_{11}}(y,x)
+\bigg( {{\alpha^\prime}_{yx}}+{{\beta^\prime}_{yx}} {\frac{{V^\prime}_{n-4}}{V_{n-4}}} \bigg) {({\omega_{n-1}})_{22}}(y,x)\nonumber\\                            -{\frac{1}{{\sqrt{r\bar r}}(1-{\frac{1}{r}})(px+qy)}}\bigg( u+v+(qx+py) {\frac{{V^\prime}_{n-4}}{V_{n-4}}} \bigg) {V_{n-1}} \bigg] 
\end{eqnarray}

The above equation reveals the reccurence relation among the partition functions of different Fibonacci generations from $(n-4)th$ to $(n+2)th$.The partition functions have entered in the above equation through the quantities given by equations from (42) to (51) .This is in conformity with the symmetry properties of the   Fibonacci chain.

\vspace{0.5cm}

\noindent     {\bf Conclusion}

\vspace{0.2cm}

   We have found an exact expression for the partition function of an Ising 
model on a reguler lattice in presence of magnetic field with open boundary conditions.This always includes closed partition functions because of the fact that
 out of four different spin configurations at the end points of the chain two configurations $\uparrow \uparrow$ and $\downarrow \downarrow$ satisfy closed boundary conditions.We have also shown that "Mirror Symmetry" is a characteristic property of all Fibonacci generations . This is the property which leads to a recurrence relation among the partition functions of different Fibonacci generations from $nth$ to $(n\pm 6)th$ . Since $nth$ and $(n\pm 6)th$ generations are topologically same , one must go through six times decimation renormalization group procedure to find scaling forms of thermodynamic functions.

\vspace{0.5cm}

  Acknowledgement: We are thankful to S.N.Karmakar of SINP,Calcutta and to S.M. Bhattacharjee of IOP,Bhubaneswar for illuminating  discussions.One ofthe authors
(SKP) thanks J.P.Chakraborty of IACS,Calcutta for lending him necessary books from his library.

\vspace{1.0cm}

[1] Yaakov Achian,T.C.Lubensky and E.W.Marshall; Phys.Rev.B33,6460(1986)

\vspace{1.0cm}

[2] Phase Transition and Critical Phenomena , article by Colin J. Thomson p177
 V1 , Edited by C. Domb and M.S.Green;  Introduction to Phase Transition and
Critical Phenomena by H. Eugene Stanley (Oxford University Press)

\vspace{1.0cm}

[3] Statistical Mechanics (Second Edition) by Kerson Huang(John Wiley  Sons)    

\vspace{1.0cm}

\end{document}